\documentstyle[aps,preprint,tighten,psfig,floats]{revtex}

\begin{document}

%\draft
\title{The Analysis of Data from Continuous Probability Distributions}
\author{Timothy E. Holy}
\address{Department of Physics, Princeton University, 
         Princeton, New Jersey, 08544}
\date{June 10, 1997}
\maketitle
\begin{abstract}
Conventional statistics begins with a model, and assigns a likelihood
of obtaining any particular set of data.  The opposite approach,
beginning with the data and assigning a likelihood to any particular
model, is explored here for the case of points drawn randomly from a
continuous probability distribution.  A scalar field theory is used to
assign a likelihood over the space of probability distributions.  The
most likely distribution may be calculated, providing an estimate of
the underlying distribution and a convenient graphical representation
of the raw data.  Fluctuations around this maximum likelihood estimate
are characterized by a robust measure of goodness-of-fit.  Its
distribution may be calculated by integrating over fluctuations.  The
resulting method of data analysis has some advantages over
conventional approaches.
\end{abstract}
%\pacs{02.50.-r}
\vskip 1cm

%%%%%%%%%%%%%%%%%%%%%%%%%%%%%%%%%%%%%%%%%%%%%%%%%%%%%%%%%%%%%%%%%%%

When the outcome of an experiment falls into
one of a few categories, the frequency of a particular outcome is an
estimate of its probability.  For example, by repeatedly flipping a
coin we learn about the probability of obtaining heads.  But when the
outcome of an experiment is one of a continuum, no finite set of data
can determine the frequency of each outcome.  One common method of
estimating the underlying probability distribution is to group
observations into categories, a procedure known as
``binning.''  The histogram (the frequency of observations in each
bin) is then used as an estimate of the underlying probability
distribution.  While binning is widely used, it has a number of
undesirable consequences.  It requires a choice of bins (both their
number and sizes), and different choices lead to different histograms.
Thus even the appearance of raw data, when presented in graphical
format, depends on arbitrary choices.  Binning also throws information
away, since different outcomes are grouped together.

An alternative approach has been presented \cite{Good_Gaskins_71,BCS}
to estimate the probability distribution.  These authors assign a
likelihood $P[Q|x_1,\ldots,x_N]$ that the distribution $Q(x)$
describes the data $x_1,\ldots,x_N$.  The underlying distribution
might then be estimated as the one which maximizes
$P[Q|x_1,\ldots,x_N]$.  By Bayes' rule,
\begin{eqnarray}
\label{eq:Bayes}
P[Q|x_1,\ldots,x_N] &=& {P[x_1,\ldots,x_N|Q] P[Q]\over
P[x_1,\ldots,x_N]} \\
&=& {Q(x_1)\cdots Q(x_N) P[Q] \over \int {\cal D}Q \, Q(x_1)\cdots
Q(x_N) P[Q]},
\end{eqnarray}
where $P[Q]$ is some {\em a priori\/} likelihood of the distribution
$Q$.  As no finite set of data can specify an arbitrary function of a
continuous variable, a choice for $P[Q]$ is necessary to regularize
the inverse problem.
This choice
encapsulates our baises in an explicit fashion.  (These biases are
implicit in other approaches, e.g., in our interpretation of a
histogram.)

What form should $P[Q]$ have?
By setting $Q(x) = \psi^2(x)$ \cite{Good_Gaskins_71},
where $\psi$ may take any
value in $(-\infty,\infty)$, we may insure that $Q$ is non-negative.
$\psi$ will be referred to as the {\em
amplitude\/} by analogy with quantum mechanics.
$P[Q]$ should incorporate our bias that $Q$ be ``smooth''
\cite{smoothnote}.
``Smoothness'' is enforced by penalizing large gradients
in $Q$---or rather, in $\psi$.  Finally, $Q$ should be
normalized.  In one dimension, the {\em a
priori\/} distribution
is
\begin{equation}
\label{eq:Ppsi}
P[\psi] = {1\over Z} \exp\left[- \int dx\, {\ell^2\over2}
(\partial_x\psi)^2 \right] \delta\left(1-\int dx\, \psi^2\right),
\end{equation}
where $Z$ is the normalization factor and $\ell$ is a constant
which controls the penalty applied to gradients.  The delta function
enforces normalization of the distribution $Q$.

The probability
$P[Q|x_1,\ldots,x_N]$ of a distribution $Q$, given the data, is
therefore
\begin{eqnarray}
\lefteqn{P[\psi|x_1,\ldots,x_N] \propto \psi^2(x_1) \cdots
\psi^2(x_N)}\qquad&&\nonumber\\ & & \times \exp\left[-\int dx\,
{\ell^2\over2} (\partial_x\psi)^2\right] \delta\left(1-\int dx\,
\psi^2\right) \\ &=& e^{-S[\psi]} \delta\left(1-\int
dx\,\psi^2\right),
\label{eq:Pofpsi}
\end{eqnarray}
where
the effective action $S$ is
\begin{equation}
S[\psi] = \int
dx\,\left({\ell^2\over2}(\partial_x\psi)^2-2\ln\psi
\sum_i\delta(x-x_i)\right).
\end{equation}

What is the most likely distribution (amplitude), given the data?  From
Eq.~(\ref{eq:Pofpsi}), this is the $\psi$ which minimizes the action, subject
to the normalization constraint.  This $\psi$ will be called
the classical amplitude, $\psi_{\rm cl}$.  To handle the normalization
constraint, we subtract a Lagrange multiplier term $\lambda(1-\int
dx\,\psi^2)$ from the action; $\psi_{\rm cl}$ satisfies the
equations
\begin{mathletters}
\label{eq:qcl}
\begin{equation}
\label{eq:qcla}
-\ell^2\partial_x^2\psi_{\rm cl}+ 2\lambda\psi_{\rm
cl}-{2\over\psi_{\rm cl}}\sum_i \delta(x-x_i) = 0,
\end{equation}
\begin{equation}
\label{eq:qclnorm}
\int dx\,\psi_{\rm cl}^2 = 1.
\end{equation}
\end{mathletters}
The solution to these equations may be written
\begin{equation}
\label{eq:qcldef}
\psi_{\rm cl}(x) = \sqrt{\kappa}\sum_ia_ie^{-\kappa|x-x_i|},
\end{equation}
where $\kappa^2=2\lambda/\ell^2$.  Each data point therefore
contributes one peak of
width $1/\kappa$ to the amplitude $\psi_{\rm cl}$.  This is reminiscent of
kernel estimation \cite{Devroye}, using the amplitude rather than the
probability distribution.
Eqs.~(\ref{eq:qcl}) imply
\begin{mathletters}
\label{eq:alam}
\begin{equation}
\label{eq:a1}
2\lambda a_i\sum_ja_je^{-\kappa|x_i-x_j|} = 1, \qquad i =
1,\ldots,N
\end{equation}
\begin{equation}
\label{eq:a2}
{N\over2\lambda} +
\sum_{i,j}a_ia_j\,\kappa|x_i-x_j|\,e^{-\kappa|x_i-x_j|} = 1.
\end{equation}
\end{mathletters}
These $N+1$ equations determine $\lambda$ and the $a_i$ as a
function of $\kappa$ \cite{numericalnote}.

Using the equation of motion, Eqs.~(\ref{eq:qcl}), the classical action
$S[\psi_{\rm cl}]$ may be written
\begin{equation}
\label{eq:Scl}
S[\psi_{\rm cl}] = N - \lambda(\kappa) - \sum_i \ln Q_{\rm cl}(x_i).
\end{equation}
For the proper choice of $\kappa$ one might hope that $Q_{\rm cl} \approx
\bar Q$, the true distribution.
Since the data points $x_i$ arise from the true distribution
$\bar Q(x)$, we expect
\begin{equation}
\label{eq:sumdelts}
\sum_i \delta(x-x_i) \approx N \bar Q(x).
\end{equation}
Therefore, the last term of Eq.~(\ref{eq:Scl}) is approximately
$N\int
dx\,\bar Q(x)\ln\bar Q(x)$, which can be interpreted as the entropy
(or the information \cite{Shannon_Weaver}).  Using perturbation theory
one may show that when
$Q_{\rm cl} \approx \bar Q$, then $\lambda \approx N$, so the first two
terms of Eq.~(\ref{eq:Scl}) (the penalty for gradients) approximately cancel
(more precisely, increase much less rapidly than $N$).

\begin{figure}
\centerline{\psfig{figure=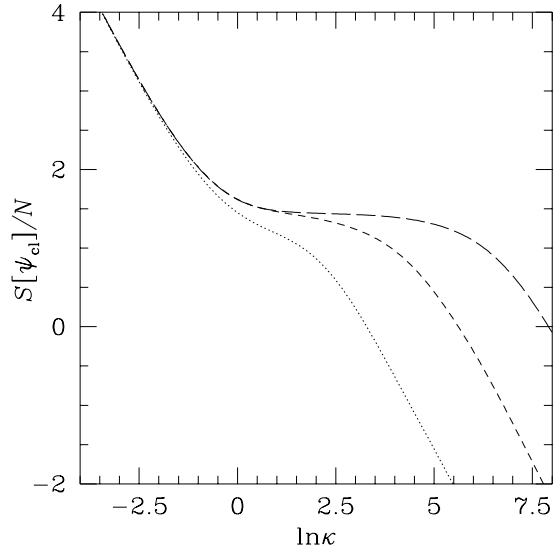,width=3in}}
\caption[]{
\label{f:scl}
The classical action, Eq.~(\ref{eq:Scl}), as a function of $\ln\kappa$
for data drawn randomly from a gaussian distribution with zero mean
and unit variance.  Long dash, $N=2000$; short dash, $N=200$; dots,
$N=20$.  }
\end{figure}

How does one choose $\kappa$?  In Figure~\ref{f:scl}, the classical
action is plotted against $\ln\kappa$ for data sets generated from a
gaussian distribution.  One sees that, over a region of width $\ln N$,
$S[\psi_{\rm cl}]$ is insensitive to the precise choice of $\kappa$.
Therefore, $\kappa$ may be chosen by finding the point of minimum
sensitivity $|dS[\psi_{\rm cl}]/d\ln\kappa|$ \cite{Stevenson,BCSnote}.

\begin{figure}
\centerline{\psfig{figure=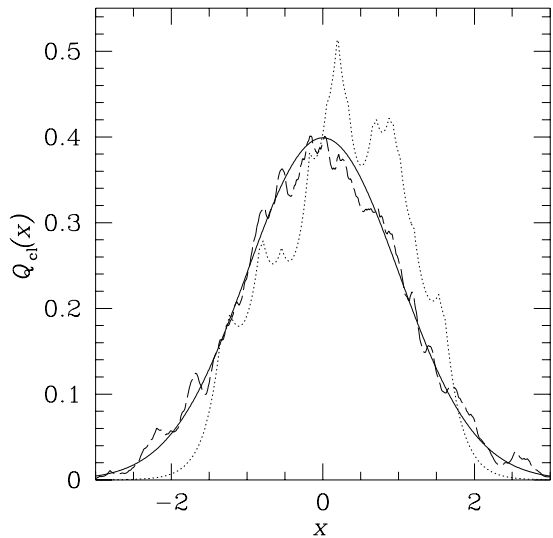,width=3in}}
\caption[]{
\label{f:gaussian}
The classical distribution $Q_{\rm cl}$, for data drawn randomly from a
gaussian distribution (solid line).  Dashed curve, $N=2000$; dotted
curve, $N=20$.
}
\end{figure}

Once $\kappa$ has been chosen, the maximum likelihood distribution
$Q_{\rm cl}(x) = \psi_{\rm cl}^2(x)$ is uniquely determined.  An example
of results from this procedure are shown in Figure~\ref{f:gaussian}.  One 
sees convergence towards the underlying distribution as $N$ increases.
Note that even for $N=20$ the estimate $Q_{\rm cl}$ is illuminating;
the advantages of this method over binning are especially great for
small data sets.

While $Q_{\rm cl}$ represents the most likely distribution, other
``nearby'' distributions should also be considered.  The action may be
expanded around the classical amplitude, which to second order in the
fluctuations $\delta\psi$ yields \cite{lambdanote}
\begin{eqnarray}
\lefteqn{S[\psi_{\rm cl}+\delta\psi] \approx S[\psi_{\rm cl}] +
{1\over4}\chi^2[\delta\psi]}\qquad&&\nonumber\\
&& + \int
dx\,\left({\ell^2\over2}(\partial_x\delta\psi)^2+\lambda\delta\psi^2\right),
\label{eq:expandS}
\end{eqnarray}
where
\begin{equation}
\label{eq:chisq}
\chi^2[\delta\psi] = 4\sum_i {\delta\psi^2(x_i)\over\psi_{\rm cl}^2(x_i)}.
\end{equation}
$\chi^2$ is a measure of the goodness of fit between a trial
distribution $Q = \psi^2$ and the data.  It is the direct analogue
of the conventional $\chi^2$ (which here will be called $\chi^2_1$); to see
this, re-write $\chi^2$ as
\begin{eqnarray}
\chi^2 &=& 4\int dx\, {(\psi(x)-\psi_{\rm cl}(x))^2\over\psi_{\rm cl}^2(x)}
\sum_i \delta(x-x_i) \nonumber\\
&\approx& 4 N\int dx\, \left(\sqrt{Q}-\sqrt{\bar Q}\right)^2
\label{eq:chisqmft}
\end{eqnarray}
using Eq.~(\ref{eq:sumdelts}).  Now suppose that $Q$ and $\bar Q$ are close,
$Q(x) = \bar Q(x) + \epsilon(x)$.  Then we may expand the difference of
square roots as
\begin{equation}
\label{eq:onchisq}
\left(\sqrt{Q}-\sqrt{\bar Q}\right)^2 \approx {1\over4}
{\epsilon^2\over\bar Q},
\end{equation}
which establishes the connection to the traditional definition
$\chi^2_1$.

This definition of $\chi^2$ has a number of advantages over
$\chi^2_1$.  Because of the quadratic dependence on $\epsilon$ and the
$\bar Q$ term in the denominator, $\chi^2_1$ is quite sensitive to the
tails of distributions.  In contrast, $\chi^2$ as defined in
Eq.~(\ref{eq:chisq}) is robust.  It is linear in $|\epsilon|$ when
$|\epsilon|$ is large, and has no potentially small term in the
denominator.  Therefore, this definition $\chi^2$ is more robust than
$\chi^2_1$.  Another advantage is that binning is unnecessary.  This
eliminates the problems of lost information and arbitrary bin-sizes
and -boundaries (and simplifies the process of fitting, as one need
not worry about shifting bin-boundaries).  Finally, this definition of
$\chi^2$ is essentially symmetric (exactly so in
Eq.~(\ref{eq:chisqmft})), and consequently is a true metric on the space
of probability distributions.  (The form in Eq.~(\ref{eq:chisqmft}) is
known as the squared Hellinger distance \cite{Devroye}.)

How is $\chi^2$ distributed?  To lowest order, the likelihood of any
particular fluctuation $\eta$ is
\begin{eqnarray}
\lefteqn{P[\eta|x_1,\ldots,x_N] \propto \delta\left(\int dx\,
\psi_{\rm cl}\eta\right)}\qquad&&\nonumber\\
&&\times \exp\left( -{1\over4}\chi^2[\eta]
- \int dx\, \left({\ell^2\over2}(\partial_x\eta)^2 +
\lambda\eta^2\right) \right).
\label{eq:Peta}
\end{eqnarray}
The distribution $P(\chi^2)$ may in principle be calculated by integrating
Eq.~(\ref{eq:Peta}) over all
$\eta$ with fixed $\chi^2$; a realizable alternative is to calculate
its Laplace transform, $\tilde P(\alpha) = \langle
e^{-\alpha\chi^2[\eta]}\rangle$, where the expectation is relative to
the distribution of $\eta$ in Eq.~(\ref{eq:Peta}).

One challenge in evaluating any integral over $\eta$ is the
``orthogonality condition'' $\delta\left(\int dx\,\psi_{\rm
cl}\eta\right)$ in Eq.~(\ref{eq:Peta}).  One way to handle this
condition is to use the delta-function representation $\delta(y) =
\lim_{\epsilon\rightarrow0^+} {1\over\sqrt{\pi\epsilon}}
e^{-y^2/\epsilon}$.  This adds a term $(\int dx\,\psi_{\rm
cl}\eta)^2/\epsilon$ to the argument of the exponential; the path
integral may then be expressed formally in terms of $\det({\bf L}+{\bf
\psi_{\rm cl}}\otimes{\bf \psi_{\rm cl}}/\epsilon)^{-1/2}$, where
${\bf L}$ is the appropriate operator (arising from the action,
Eq.~(\ref{eq:expandS})) and ${\bf \psi_{\rm cl}}\otimes{\bf \psi_{\rm
cl}}$ is the matrix with the $(x,x')$ element equal to $\psi_{\rm
cl}(x)\psi_{\rm cl}(x')$.  The non-local terms proportional to
$1\over\epsilon$ are large and must be handled first.  We know that
$\lim_{\epsilon\rightarrow0^+} \epsilon\det({\bf L}+{\bf \psi_{\rm
cl}}\otimes{\bf \psi_{\rm cl}}/\epsilon)$ must be finite, so all the
terms diverging worse than ${1\over\epsilon}$ in the determinant must
vanish. (This happens because of the all-order singularity of the
matrix ${\bf \psi_{\rm cl}}\otimes{\bf \psi_{\rm cl}}$.)  So even
though ${1\over\epsilon}$ is large, we may evaluate this determinant
exactly by working to first order in ${1\over\epsilon}$.  Therefore
\begin{eqnarray}
\det\left({\bf L}+{{\bf \psi_{\rm cl}}\otimes{\bf \psi_{\rm
cl}}\over\epsilon}\right) &=& \det{\bf L}\det\left(1+{{\bf L}^{-1}{\bf
\psi_{\rm cl}}\otimes{\bf \psi_{\rm
cl}}\over\epsilon}\right)\nonumber\\ & =& \det{\bf L}\,\left(1 +{{\rm
Tr}({\bf L}^{-1}{\bf \psi_{\rm cl}}\otimes{\bf \psi_{\rm
cl}})\over\epsilon}\right).
\end{eqnarray}
Now we can take the limit $\epsilon\rightarrow0^+$;
the integral over all $\eta$ is now complete.
The distribution of $\chi^2$ (properly
normalized) is therefore
\begin{equation}
\tilde P(\alpha) = \left[{D(\gamma) T(\gamma)\over D(1)
T(1)}\right]^{-1/2},
\label{eq:Ptil}
\end{equation}
where $\gamma = 4\alpha+1$,
\begin{equation}
\label{eq:D}
D(\gamma) = {\det(-\ell^2\partial_x^2 + 2\lambda +
2\gamma\sum_i\delta(x-x_i)/Q_{\rm cl})\over\det(-\ell^2\partial_x^2 +
2\lambda)},
\end{equation}
\begin{equation}
\label{eq:T}
T(\gamma) = \int dx\,dx'\,K_\gamma(x,x')\psi_{\rm cl}(x)\psi_{\rm cl}(x'),
\end{equation}
and the propagator $K_\gamma =
{\bf L}^{-1}$ satisfies
\begin{equation}
\label{eq:K}
-\ell^2\partial_x^2K_\gamma + 2\lambda K_\gamma+{2\gamma\over Q_{\rm
cl}}\sum_i \delta(x-x_i)K_\gamma = \delta(x-x').
\end{equation}

The terms of Eq.~(\ref{eq:Ptil}) can be evaluated exactly.  First,
consider the ratio
of the determinants, Eq.~(\ref{eq:D}).
Standard techniques \cite{Coleman} allow one to express
$D(\gamma)$
as the limit as $x\rightarrow\infty$ of the function $E(x;\gamma)$, where $E$
satisfies
\begin{equation}
\label{eq:dr}
-\partial_x^2 E - 2\kappa \partial_x E +
{\gamma\kappa^2\over\lambda Q_{\rm cl}}\sum_i \delta(x-x_i) E = 0
\end{equation}
and $E(x) = 1$ for $x$ smaller than the smallest data point.  Between
data points, $E(x) = E_i + F_i
e^{-2\kappa (x-x_i)}$,
and a short calculation shows that $E_i$ and $F_i$ satisfy a simple
recursion relation.

The traces $T(\gamma)$ are computed as follows: let $g_\gamma(x) =
\int dx'\,K_\gamma(x,x')\psi_{\rm cl}(x')$ and $g_0 = \int
dx'\,K_0(x,x')\psi_{\rm cl}(x')$.  $g_\gamma$ may be parametrized as
\begin{equation}
g_\gamma(x) = g_0(x) + {\sqrt{\kappa}\over 4\lambda} \sum_i c_i
e^{-\kappa|x-x_i|},
\end{equation}
and from Eq.~(\ref{eq:K}) the $c_i$ satisfy the linear equations
\begin{equation}
\label{eq:tracecoef}
c_i + \gamma \mu_i\sum_j\left[c_j+(1+\kappa|x_i-x_j|)
a_j\right]e^{-\kappa|x_i-x_j|} = 0.
\end{equation}
where $\mu_i = {\kappa\over 2\lambda Q_{\rm cl}(x_i)}$.
Then $T(\gamma)$ may be expressed in terms of the $c_i$ by computing
the remaining integral over $x$ (which may be done analytically).

This completes the evaluation of the distribution of $\chi^2$.  One
sees that different data sets yield different
$P(\chi^2)$.  Therefore, it may be illustrative to consider the limit
of large $N$, where the distribution of $\chi^2$ assumes a more
universal form.

In the limit of large $N$, we may put $Q_{\rm cl}\approx\bar Q$ and
$\lambda\approx N$.
We write $\chi^2$ in a form similar to
Eq.~(\ref{eq:chisqmft}), but introduce a small but necessary change:
$\chi^2 \approx 4 N \int_{\bf X} dx\, \delta\psi^2$
where, heuristically, ${\bf X}$ is the region over which we may expect to
find data
points.  We need only the size $X$ of ${\bf X}$, which may be defined as
$X = {1\over N}\sum_i {1\over Q_{\rm cl}(x_i)}$.
The determinant operator is
$\ell^2(-\partial_x^2 + \kappa^2)$ outside ${\bf X}$, and
$\ell^2(-\partial_x^2 + \kappa^2(1+\gamma))$ inside ${\bf X}$.
Then the ratio of determinants (ignoring all but the exponential-order
terms) is
$
D(\gamma) \approx e^{\kappa\left(\sqrt{1+\gamma} - 1\right)X}.
$
The traces do not contribute to the exponential-order terms.  Consequently,
\begin{equation}
\label{eq:PtilN}
\tilde P(\alpha) \approx e^{-\langle\chi^2\rangle
\left(\sqrt{1+2\alpha}-1\right)},
\end{equation}
where $\langle\chi^2\rangle\approx\kappa X/\sqrt{2}$.
Note that if
we identify
$1/\kappa$ as the effective bin width, then $\langle\chi^2\rangle$ is
approximately $1/\sqrt{2}$ per bin, i.e., $\approx 0.7$ per degree of
freedom.
We may invert the Laplace transform in Eq.~(\ref{eq:PtilN}) to obtain
\begin{equation}
P(z) \approx {\langle\chi^2\rangle\over\sqrt{2\pi z^3}}
\exp\left[\langle\chi^2\rangle\left(1 - {z\over2\langle\chi^2\rangle}
- {\langle\chi^2\rangle\over2z}\right)\right].
\end{equation}

The conventional approach to statistics emphasizes the model: given a
model, one calculates the likelihood of obtaining a particular data
set.  This likelihood is measured by the conventional $\chi^2$.  Its
distribution is over (hypothetical) repeated trials of the experiment,
assuming gaussian errors.  In contrast, the approach presented here
emphasizes the data: given a data set, one calculates the likelihood
that it is described by a particular model.  This likelihood is
measured by $\chi^2$; its distribution is over all possible models.

The approach presented here has two major advantages over conventional
methods.  First, it provides a technique for visualizing data sets,
retaining all the information in the data and requiring no arbitrary
choices.  Second, it provides a robust measure of goodness-of-fit.
Its distribution can be calculated, and so may be used for statistical
analysis.  The availability of a fast algorithm \cite{numericalnote}
makes computation time negligible even for large data sets.  This
technique should be generalizable to higher
dimensions \cite{BCS}.

\acknowledgements
TEH is supported by a Lucent Technologies
Ph.D. Fellowship.  I thank S. Strong and W. Bialek for useful
conversations.  This work is dedicated to W.  H. Press,
B. P. Flannery, S. A. Teukolsky, and W.  T. Vetterling.

%\bibliographystyle{prsty}
%\bibliography{prlsubmit}

\begin{thebibliography}{10}

\bibitem{Good_Gaskins_71}
I.~J. Good and R.~A. Gaskins, Biometrika {\bf 58},  255  (1971).

\bibitem{BCS}
W. Bialek, C.~G. Callan, and S.~P. Strong, Phys.\ Rev.\ Lett. {\bf 77},  4693
  (1996).

\bibitem{smoothnote}
Without such a bias, e.g., if we choose $P[Q] = 1$, the most likely $Q$ is the
  solipsistic ${1\over N} \sum_i \delta(x-x_i)$ \cite{BCS}.

\bibitem{Devroye}
L. Devroye, {\em A Course in Density Estimation} (Birkh\"auser, Boston, 1987).

\bibitem{numericalnote}
Eqs.~(\ref{eq:alam}) are solved by Newton's method, i.e., by linearizing around
  the solution. An $N\times N$ block of the resulting matrix equation may be
  put in the form $\bf A u = b$, where $\bf A = 1 + \Delta W$, $\bf\Delta$ is a
  diagonal matrix, and ${\bf W}_{ij} = e^{-\kappa|x_i-x_j|}$. Note that
  Eq.~(\ref{eq:tracecoef}) has the same form. Solving this linear equation is
  nominally an ${\rm O}(N^3)$ process. However, it is possible to do much
  better, because (when $x_1,\ldots,x_N$ are sorted in increasing order)
  ${\bf\Omega} = {\bf W}^{-1}$ is tridiagonal. Using ${\bf\Omega}^{-1}$ in
  place of ${\bf W}$ allows all operations to be performed in ${\rm O}(N)$
  time, a very significant savings for large data sets. Source code may be
  requested from holy{@}puhep1.princeton.edu. Computational issues were also
  considered in J. Ghorai and H. Rubin, J. Stat.\ Comput.\ Simul. {\bf 10}, 65
  (1979). Existence and uniqueness of a non-negative $\psi_{\rm cl}$ was shown
  in G.~F. de~Montricher, R.~A. Tapia, and J.~R. Thompson, Ann.\ Stat. {\bf 3},
  1329 (1975).

\bibitem{Shannon_Weaver}
C.~E. Shannon and W. Weaver, {\em The Mathematical Theory of Communication}
  (University of Illinois Press, Urbana, 1949).

\bibitem{Stevenson}
P.~M. Stevenson, Phys.\ Rev.\ D {\bf 23},  2916  (1981).

\bibitem{BCSnote}
In Ref.~\cite{BCS}, the smoothing parameter cannot be set until the expected
  value $\langle Q(x_1)\cdots Q(x_N)\rangle$ has been calculated, which
  requires integrating over the fluctuations and a WKB analysis. Here the
  fluctuations ($[2\lambda D(1)T(1)]^{-1/2}$) do not qualitatively change
  Figure~\ref{f:scl}; even the optimum choice for $\kappa$ is changed little.
  Note that the choice $\ell_*$ in Ref.~\cite{BCS} is (regrettably) zero for
  many common distributions $\bar Q$.

\bibitem{lambdanote}
One must decide whether the $\lambda$ terms are included in computing the
  fluctuations. The two choices yield very similar results; the version used
  here turns out to be somewhat simpler to implement.

\bibitem{Coleman}
S. Coleman,  in {\em Aspects of Symmetry} (Cambridge University Press,
  Cambridge, 1975), Chap.~7 (Appendix 1).

\end{thebibliography}

\end{document}